\documentclass[preprint2]{aastex}

\newcommand{\neii}{[Ne~{\sc ii}]}
\newcommand{\hi}{H~{\sc i}}
\newcommand{\neiii}{[Ne~{\sc iii}]}
\newcommand{\hii}{H~{\sc ii}}
\newcommand{\civ}{C~{\sc iv}}
\newcommand{\ovi}{O~{\sc vi}}
\newcommand{\oi}{[O~{\sc i}]}
\newcommand{\caii}{Ca~{\sc ii}}

\shorttitle{Detection of [Ne~{\sc ii}] emission}
\shortauthors{Pascucci et al.}

\begin{document}


\title{Detection of [Ne~{\sc ii}] Emission from Young Circumstellar Disks}

%====== Other possible titles
% 1) Detection of [Ne~{\sc ii}] emission lines:  tracing the hot disk atmosphere of young circumstellar disks


\author{I. Pascucci\altaffilmark{1}, D. Hollenbach\altaffilmark{2}, J. Najita\altaffilmark{3},
J. Muzerolle\altaffilmark{1}, U. Gorti\altaffilmark{4}, G. J. Herczeg\altaffilmark{5},
L. A. Hillenbrand\altaffilmark{5}, J. S. Kim\altaffilmark{1},  J. M. Carpenter\altaffilmark{5}
M. R. Meyer\altaffilmark{1},  E. E. Mamajek\altaffilmark{6}, J. Bouwman\altaffilmark{7}} 
%D. M. Watson\altaffilmark{8}}
%\affil{Steward Observatory, The University of Arizona, Tucson, AZ 85721.}


\altaffiltext{1}{Steward Observatory, The University of Arizona, Tucson, AZ 85721.}
\altaffiltext{2}{University of California, Berkeley, CA 94720}
\altaffiltext{3}{National Optical Astronomy Observatory, Tucson, AZ 85719}
\altaffiltext{4}{NASA Ames Research Center, Moffett Field, CA 94035}
\altaffiltext{5}{California Institute of Technology, Pasadena, CA 91125}
\altaffiltext{6}{Harvard--Smithsonian Center for Astrophysics, Cambridge, MA 02138}
\altaffiltext{7}{Max Planck Institute for Astronomy, Heidelberg, Germany}
%\altaffiltext{8}{Department of Physics and Astronomy, University of Rochester, Rochester, NY 14627}


\begin{abstract}
We report the detection of  \neii{} emission at 12.81\,\micron{} in four out of the six
optically thick dust disks observed as part of the FEPS {\it Spitzer} Legacy 
program. In addition, we detect a \hi (7-6) emission line at 12.37\,\micron{} from the source
RX~J1852.3-3700. Detections of \neii{} lines are favored by low mid--infrared
excess emission.  
%We find that \neii{} line luminosities correlate with stellar X--ray luminosities and
%anti--correlate with mass accretion rates. 
Both stellar X--rays and  extreme UV (EUV) photons can sufficiently ionize the disk surface
to reproduce the observed line fluxes, suggesting that emission from Ne$^+$ originates in
the hot disk atmosphere. On the other hand, the \hi (7-6) line is not associated with the
gas in the disk surface and magnetospheric accretion flows can account for at most $\sim$30\% of the observed flux. We conclude that accretion shock regions and/or the stellar corona could contribute to most of the \hi (7-6) emission.
Finally, we discuss the observations necessary to identify whether stellar X--rays or EUV photons are the dominant ionization mechanism for Ne atoms. Because the observed \neii{} emission probes very small amounts of gas in the disk surface ($\sim 10^{-6}\,M_{\rm J}$)
we suggest using this gas line to determine the presence or absence of gas in more evolved circumstellar disks.

\end{abstract}





\section{Introduction}
Young stars are often surrounded by gas and dust disks that may succeed in forming planets.
The properties and evolution of their dust and gas components  are key to understanding planet formation and 
the diversity of extrasolar planetary systems. 

Circumstellar dust has been extensively studied in young and old disks since the IRAS mission. 
Grain growth  has been identified in disks around stars with masses ranging from a few solar masses \citep{bouwman01} down to the brown dwarf regime \citep{apai05} suggesting that the first steps of  planet formation are ubiquitous.
Detailed mineralogy  reveal chemical processing similar to those that occurred in the early solar system
(see  \citealt{natta07} for a review on dust in protoplanetary disks; \citealt{CMalexander07} and \citealt{wooden07} for dust processing in the solar nebula and in disks). 
On the other hand, the properties and evolution of  circumstellar gas are less well characterized.

Filling this gap is one of the goals of the Formation and Evolution of Planetary Systems (FEPS) 
{\it Spitzer} legacy program \citep{meyer06}.  In two previous contributions we derived gas mass upper limits for a sample of
16 sun-like stars surrounded by optically thin dust disks and discussed implications for the formation of terrestrial,
giant, and icy planets \citep{holl05,pascucci06}. 

Here, we present high-resolution {\it Spitzer}  spectra for the six FEPS 
targets with excess emission beginning at or shortward of the 8\,\micron{} IRAC band. Their IRAC colors are consistent with those of accreting classical T~Tauri stars \citep{silverstone06} suggesting that these stars are surrounded by optically thick dust disks. We report the detection of the \hi{} (7--6) line  at 12.37\,\micron{} in one system and of the \neii{} line at 12.81\,\micron{} in four systems (Sect.~\ref{S:detections}).
Because Ne atoms have a large ionization potential (21.6\,eV), the detection of \neii{} lines  is of particular interest to 
assess the role of stellar X--ray and EUV ($h\nu > 13.6$\,eV) photons on the disk chemistry and to explore the conditions for disk photoevaporation \citep{glass07,gorti07}.
We discuss in detail predictions from the proposed X--ray and EUV models and future observations that will be able to 
identify the dominant ionization mechanism (Sect.~\ref{S:origin} and Sect.~\ref{S:discussion}). 
Since both models find that the \neii{} line probes  small amounts 
of gas on the surface of circumstellar disks, we also suggest using this tracer to place stringent constraints on the presence or absence of gas in more evolved disks.

\begin{deluxetable}{cl c cccc cc cc}
\tabletypesize{\scriptsize}
\rotate
\tablecaption{Main properties of the targeted optically thick dust disks.  \label{T:Stars}}
%Additional information like infrared fluxes are available in Table~2 from \citet{silverstone06} and Table~1 from \citet{bouwman06}. \label{T:Stars}}
\tablewidth{0pt}
\tablehead{
\colhead{ID\#}&\colhead{Source}&\colhead{2MASS~J\tablenotemark{a}}&\colhead{SpTy}&\colhead{$d$}&\colhead{Age}& \colhead{Ref.} &
\colhead{$T_{\rm eff}$}&\colhead{$A_{\rm v}$}&\colhead{log($L_\star$)}&\colhead{log($L_{\rm x}$)\tablenotemark{b}} \\
\colhead{} & \colhead{} & \colhead{} &  \colhead{} & \colhead{[pc]} & \colhead{[Myr]}& \colhead{} &
\colhead{[K]} &  \colhead{[mag]} &  \colhead{[L$_\sun$]}  & \colhead{[erg/s]}   }
\startdata
1 & RX~J1111.7-7620        & 11114632-7620092  &K1   & 163$\pm$10 & 5 &  1,2,3& 4621& 1.30 & 0.27 &30.56$\pm$0.16  	 \\ 
2 & PDS~66 		   & 13220753-6938121  &K1   & 86$\pm$7  & 17&  4   & 5228           & 1.22  & 0.10 &30.20$\pm$0.09 	 \\ 
3 & HD~143006              & 15583692-2257153  &G6/8 & 145$\pm$10 & 5 &  5,6,7& 5884  & 1.63 & 0.39 &30.14$\pm$0.17  		\\ 
4 & [PZ99]~J161411.0-230536& 16141107-2305362  &K0   & 145$\pm$10 & 5 &  8,6,7& 4963& 1.48 &0.50 &30.54$\pm$0.10  	  \\
5 & RX~J1842.9-3532        & 18425797-3532427  &K2   & 140$\pm$10 & 4 &  9,10,3& 4995   & 1.03 &0.05 & 30.37$\pm$0.18  		 \\
6 & RX~J1852.3-3700        & 18521730-3700119  &K3   & 140$\pm$10 & 4 &  9,10,3& 4759   & 0.92 &-0.17 &30.57$\pm$0.13  		 \\
%%%%%%%%%%% Old ordering by IR excess emission
%1 & RX~J1852.3-3700        & 18521730-3700119  &K3   & 140 & 4 &  1,2,3& 4759 & 0.92 & 30.57$\pm$0.13  		 \\
%2 & RX~J1842.9-3532        & 18425797-3532427  &K2   & 140 & 4 &  1,2,3& 4995 & 1.03 & 30.37$\pm$0.18  		 \\
%3 & RX~J1111.7-7620        & 11114632-7620092  &K1   & 163 & 5 &  4,5,3& 4621 & 1.30 & 30.73$\pm$0.10  	 \\ 
%4 & [PZ99]~J161411.0-230536& 16141107-2305362  &K0   & 145 & 5 &  6,7,8& 4963 & 1.48 & 30.54$\pm$0.10  	  \\
%5 & HD~143006              & 15583692-2257153  &G6/8 & 145 & 5 &  9,7,8& 5884 & 1.63 & 30.14$\pm$0.17  		\\ 
%6 & PDS~66 		   & 13220753-6938121  &K1   & 86  & 17&  10   & 5228 & 1.22  & 30.20$\pm$0.09 	 \\ 
\enddata
%\tablerefs{
%(1) \citet{neu00}
%}
\tablenotetext{a}{The 2MASS source name includes the J2000 sexagesimal, equatorial position in the form: hhmmssss+ddmmsss 
\citep{cutri03}.}
\tablenotetext{b}{X--ray luminosities in the ROSAT 0.1--2.4\,keV energy band, see Sect.~\ref{S:Obs} for details.} 
%\citep{alcala97,voges99,voges00} following
%\citet{fleming95}, and adopting the distances in the table. We assumed a 10\,pc error on the distance for all stars except PDS~66, for which
%\citet{mamajek02} quote an uncertainty of 7\,pc.
% Our X--ray luminosities agree with values from the literature 
%(\citealt{alcala97} for source 1; \citealt{mamajek02} for source 2; 
%\citealt{sciortino98} for source 3; \citealt{neuhauser00} for sources 5 and 6). 
%Note that the intrinsic variability of log($L_{\rm x}$) due to stellar activity is larger than the quoted uncertainty and amounts to at least  a few tenths of a dex (e.g. \citealt{marino03}). 
%Because source 4 appears extended in the ROSAT PSPC image, we searched for an independent measurement of its X--ray flux. Its count--rates from the XMM--Newton MOS1 and MOS2 cameras in the 0.2--2\,keV bandpass convert to log($L_{\rm x}$)=30.8\, erg/s (XMM--Newton Serendipitous Source Catalogue, 1XMM, http://xmmssc-www.star.le.ac.uk/). Since an increase of 0.26\,dex in luminosity could be simply due to stellar activity,  we prefer to adopt the luminosity from ROSAT for consistency with the other sources.
%}
\tablecomments{Spectral types are from optical spectroscopy with accuracy of 1 subtype. The stellar colors B--V and V--K are used in conjunction with the spectral type to estimate effective temperatures ($T_{\rm eff}$).
 Visual extinctions ($A_{\rm v}$) are computed from the spectral type and stellar colors. We refer to Carpenter et al. 2007 in preparation for  details.
 Stellar luminosities are  computed from the best fit Kurucz stellar models to optical and near--infrared observations of the stellar photosphere (Carpenter et al. 2007 in prep.) and the distances in this table. 
 }
\tablerefs{
(1) \citealt{alcala95}; (2) \citealt{luhman07};
(3) Hillenbrand et al.~2007 in prep.; (4) \citealt{mamajek02};
(5) \citealt{houk88}; (6) \citealt{dezeeuw99}; (7) \citealt{preibisch02};
(8) \citealt{preibisch98}; 
(9) \citealt{neuhauser00}; (10) \citealt{neuhauser07}
%(1) \citealt{neuhauser00}; (2) \citealt{neuhauser07}; (3) Hillenbrand et al.~2007 in prep.;
%(4) \citealt{alcala95}; (5) \citealt{luhman07};
%(6) \citealt{preibisch98}; (7) \citealt{dezeeuw99}; (8) \citealt{preibisch02}; 
%(9) \citealt{houk88}
%(10) \citealt{mamajek02}
}
\end{deluxetable}

\section{Observations and Data Reduction}\label{S:Obs}
The general properties of the 328 stars in the FEPS sample are described in \citet{meyer06}.
We summarize in Table~\ref{T:Stars} the main properties of the six FEPS sources surrounded by optically thick dust disks. In Cols.~4, 5, and 6 we give the star spectral types (SpTy), distances ($d$), and ages (Age). References for these quantities are provided in Col.~7. The stellar effective temperatures ($T_{\rm eff}$), visual extinctions ($A_{\rm V}$), and bolometric luminosities ($L_\star$) are listed in Cols.~8, 9, and 10. The last column summarizes the source X-ray luminosities ($L_{\rm X}$).

To compute the X--ray luminosities we used the ROSAT PSPC All--Sky Survey
count--rates and HR1 hardness ratios \citep{alcala97,voges99,voges00} following
\citet{fleming95}, and adopting the distances in Table~\ref{T:Stars}. These X-ray
luminosities are representative for the energy band 0.1--2.4\,keV (or 120--5\,\AA).
The errors in  $L_{\rm X}$ include the uncertainties in the count--rates, HR1, and
distances. Note however that the intrinsic variability of log($L_{\rm x}$) due to stellar activity is larger than the quoted uncertainty and amounts to at least  a few tenths of a dex (e.g. \citealt{marino03}). 
 Our X--ray luminosities agree with values from the literature 
(\citealt{alcala97} for source 1; \citealt{mamajek02} for source 2; 
\citealt{sciortino98} for source 3; \citealt{neuhauser00} for sources 5 and 6). 
Because source 4 appears extended in the ROSAT PSPC image, we also searched for an independent measurement of its X--ray flux. Its count--rates from the XMM--Newton\footnote{XMM--Newton Serendipitous Source Catalogue, 1XMM, http://xmmssc-www.star.le.ac.uk/} MOS1 and MOS2 cameras in the 0.2--2\,keV bandpass convert to log($L_{\rm x}$)=30.8\, erg/s. Since an increase of 0.26\,dex in luminosity could be simply due to stellar activity,  we prefer to adopt the luminosity from ROSAT for consistency with the other sources.

Among the sources listed in Table~\ref{T:Stars}, HD~143006 is the only one that was included in the FEPS program as part of the gas detection experiment being a known dust disk from IRAS \citep{sylvester96}.
The other five dust disks have been recently identified by our group from IRAC/{\it Spitzer} photometry
\citep{silverstone06}.  The spectral energy distributions of these five systems, including photometry from IRAC and MIPS as well as IRS low--resolution spectra, are presented in \citet{silverstone06}.  Hillenbrand et al. in prep. show that single temperature blackbody fits to the  33 and 70\,\micron{} data cannot account for the excess emission at shorter wavelengths in any of the six sources in Table~\ref{T:Stars}, indicating the presence of warm inner disk material. These optically thick dust disks are also the only six FEPS sources exhibiting dust features in the {\it Spitzer}/IRS low--resolution spectra. 
\citet{bouwman07} present a detailed analysis of their mineralogy and find that 
 RX~J1842.9-3532 is surrounded by an almost primordial disk (flared geometry and 1\,\micron --sized grains) while [PZ99]~ J161411.0-230536 has the most processed disk (close to flat geometry and large 5\,\micron{}  grains).

In Sect.~\ref{S:ReduIRS} we describe the observations and data reduction of the high--resolution infrared spectra.
We complemented the {\it Spitzer} data with optical spectra (Sect.~\ref{S:ReduOPT}) that are used to
estimate mass accretion rates  (Sect.~\ref{S:Mdot}). 

\subsection{High--resolution IRS Spectra}\label{S:ReduIRS}
{\it Spitzer}/IRS high--resolution spectra for  the 6 FEPS optically thick dust disks were
obtained between August 2004 and September 2005. Observations were done in the Fixed
Cluster--Offsets mode with two nod positions on--source (located at  1/3 and 2/3 of the slit
length) and two additional sky measurements (1$'$ east of the nod1 and nod2 positions) 
acquired just after the on--source exposures.  We used these sky exposures to remove the
infrared background and reduce the number of rogue pixels as described in
\citet{pascucci06}.  The PCRS or the IRS Peak--up options were used to place and hold the 
targets in the spectrograph slit with positional uncertainties always better than 1\arcsec{}
(1 sigma  radial).  Exposure times were chosen to detect  a 5\% line--to--continuum ratio with
a signal--to--noise of 5 (see Table~\ref{T:Obs}). In the following we focus on the reduction
and analysis of the SH module (9.9--19.6\,\micron) where we detected gas lines in four out of
six targets. No gas lines are detected in the wavelength range 18.7--37.2\,\micron{} that is covered by the LH module.
The SH module is a cross--dispersed echelle spectrograph with a resolving power 
of $\sim 650$ in the spectral range from 9.9--19.6\,\micron , corresponding to a spectral resolution of 
0.015\,\micron{} around 10\,\micron . The detector has a plate scale of 2.3\arcsec /pixel and the slit aperture
has a size of $2\times5$ pixels, thus covering a region of $\sim 1600\times 700$\,AU around a star at 140\,pc.

Data reduction was carried out as in \citet{pascucci06} starting from the Spitzer Science Center (SSC)  S13.2.0 {\it droop} products. We fixed pixels marked bad in the ÓbmaskÓ files with flag value equal to
2$^9$ or larger, thus including anomalous pixels due to cosmic--ray saturation early  in the
integration, or preflagged as unresponsive. We also inspected visually all the SH exposures
to catch additional rogue pixels and found less than 5 per frame. These bad and rogue pixels
were corrected using the SSC {\it irsclean} package as explained in \citet{pascucci06}.

We flux calibrated the extracted spectra using nine independent observations (over four
different {\it Spitzer} campaigns, from C21 to C24)  of the bright standard star
$\xi$~Dra and the MARCS stellar atmosphere model degraded to the spectrograph's resolution
and sampling\footnote{http://ssc.spitzer.caltech.edu/irs/calib/templ/}. The dispersion in
the mean fluxes of these calibrators is always less than 1.5\% between 10 and 20\,\micron.  
We verified the absolute flux calibration by comparing the high--resolution spectra to the low--resolution spectra from
\citet{bouwman07} at a wavelength in between the \neii{} and the \hi (7-6)
lines. We found deviations  less than $\sim$10\% at 12.6$\pm0.1$\,\micron{},
that are within the absolute photometric accuracy estimated by the SSC,  for all objects
except for  RX~J1111.7-7620. For this source the SH spectrum has 30\% higher flux but maintains
the same shape as the low--resolution spectrum. We verified that this higher flux is simply
due to the non--zero background emission in the vicinity of the source (the sky position we
acquired for the SH spectrum is too far away from the source to be representative of the
nearby emission). We have also tested that the low--resolution spectra agree
with the IRAC photometry at 8\,\micron{} and the MIPS photometry at
24\,\micron{}\footnote{http://ssc.spitzer.caltech.edu/legacy/fepshistory.html}. Thus, to
improve our flux estimates and upper limits we scaled the high--resolution spectra to the
flux of the low-resolution spectra at 12.6\,\micron .
%12.6($\pm$0.1)\,\micron 

\begin{deluxetable}{l ccccc}
\tabletypesize{\scriptsize}
\tablewidth{0pt}
\tablecaption{Log of the IRS short-wavelength high-resolution observations.\label{T:Obs}}
\tablehead{
\colhead{Source}&\colhead{AOR Key}&\colhead{SH}&\colhead{Peak-up}  \\
\colhead{} & \colhead{} & \colhead{time$\times$ncycles} &  \colhead{mode} }
\startdata
RX~J1111.7-7620              & 5451776 &  31.5$\times$12 &  PCRS	       \\  % DATE_OBS= '2005-06-02T03:44:47.663'
PDS~66	     & 5451264 & 6.3$\times$4      & PCRS       \\   % DATE_OBS= '2005-04-17T14:09:32.348'
HD~143006                    & 9777152 & 6.3$\times$18    &  PCRS	      \\  % DATE_OBS= '2004-08-08T05:20:03.208'
$[$PZ99$]$~J161411.0-230536      & 5453824 & 31.5$\times$4    & PCRS      \\  % DATE_OBS= '2005-08-10T13:38:24.688'
RX~J1842.9-3532              & 5451521 &  31.5$\times$10 &  IRS-red	       \\  % DATE_OBS= '2005-09-14T03:48:34.106'
RX~J1852.3-3700              & 5452033 & 121.9$\times$4 & PCRS	       \\  % DATE_OBS= '2005-09-14T05:25:25.505'
%% Note that there is an error in one of the cycle for PDS66!!!
\enddata
\end{deluxetable}

\subsection{Optical Spectra}\label{S:ReduOPT}
Optical spectra were obtained in several observing runs and with different telescopes.
The spectra of HD~143006, RX~J1842.9-3532, and RX~J1852.3-3700 were acquired in July 2001 and June 2003 with the Palomar 60--inch
telescope and facility spectrograph in its echelle mode \citep{McCarthy88}.  A 1\farcs43$\times$7\farcs36 slit was used in combination with a (2--pixel) resolving power of $\sim$19,000.
Details on the observational strategy and data reduction are provided in Sects.~2 and 3 of \citet{white06}.  \citet{white06} also 
derive several stellar properties from the Palomar 60--inch spectra for these and other FEPS sources, including radial and rotational velocities, Li~{\sc i} $\lambda$ 6708 and H$\alpha$ equivalent widths, chromospheric activity index $R'_{HK}$, and temperature-- and gravity--sensitive line ratios.

The spectrum of [PZ99]~J161411.0-230536 was acquired  during
commissioning of the East Arm Echelle spectrograph on the Hale 200--inch telescope at Palomar.
The on--source exposure time was 2,000 seconds. The spectrum covers most of the 4000--9350\,\AA{}
region with a resolution that varies from 16,000 to 27,000 within each order.  We extracted the
spectrum and corrected for scattered light using custom routines in IDL, following the main
steps described in  \citet{white06}.

The spectrum of RX~J1111.7-7620 was acquired in March 2003 with the MIKE echelle spectrograph on
the Magellan Clay 6.5--m telescope \citep{bernstein02}. The star was
observed in a 360--second exposure with the MIKE Red CCD in the standard
setup with the 0.35\arcsec{} slit, and 2--pixel resolution of R\,$\simeq$\,36,000
covering the wavelength range 4800--8940\,\AA. The data were reduced using
the MIKE Redux IDL package\footnote{http://web.mit.edu/~burles/www/MIKE}.

Finally, the spectrum of PDS~66 was acquired in April 2002 with the echelle spectrograph on the CTIO 4 meter Blanco telescope. We used the 31.6 red long echelle grating covering the wavelength range between 3,000--10,000\,\AA{} and  a 0\farcs8$\times$3\farcs3 slit with a 2--pixel resolution of R\,$\simeq$45,000. The exposure time on--source was 120 seconds.  For the data reduction we used the IRAF packages {\it quadred}/{\it echelle} that can treat  multiamplifier echelle data.  
%Hbeta 0.486nm

\section{Results}
In Sect.~\ref{S:detections} we present the gas line detections from 
the {\it Spitzer} spectra of the six FEPS optically thick dust disks. 
We also compute mass accretion
rates from the Balmer emission profiles (Sect.~\ref{S:Mdot}) and use them in Sect.~\ref{S:origin}
to search for correlations with the flux of the detected infrared lines.

\begin{figure*}
\includegraphics[angle=90,scale=0.7]{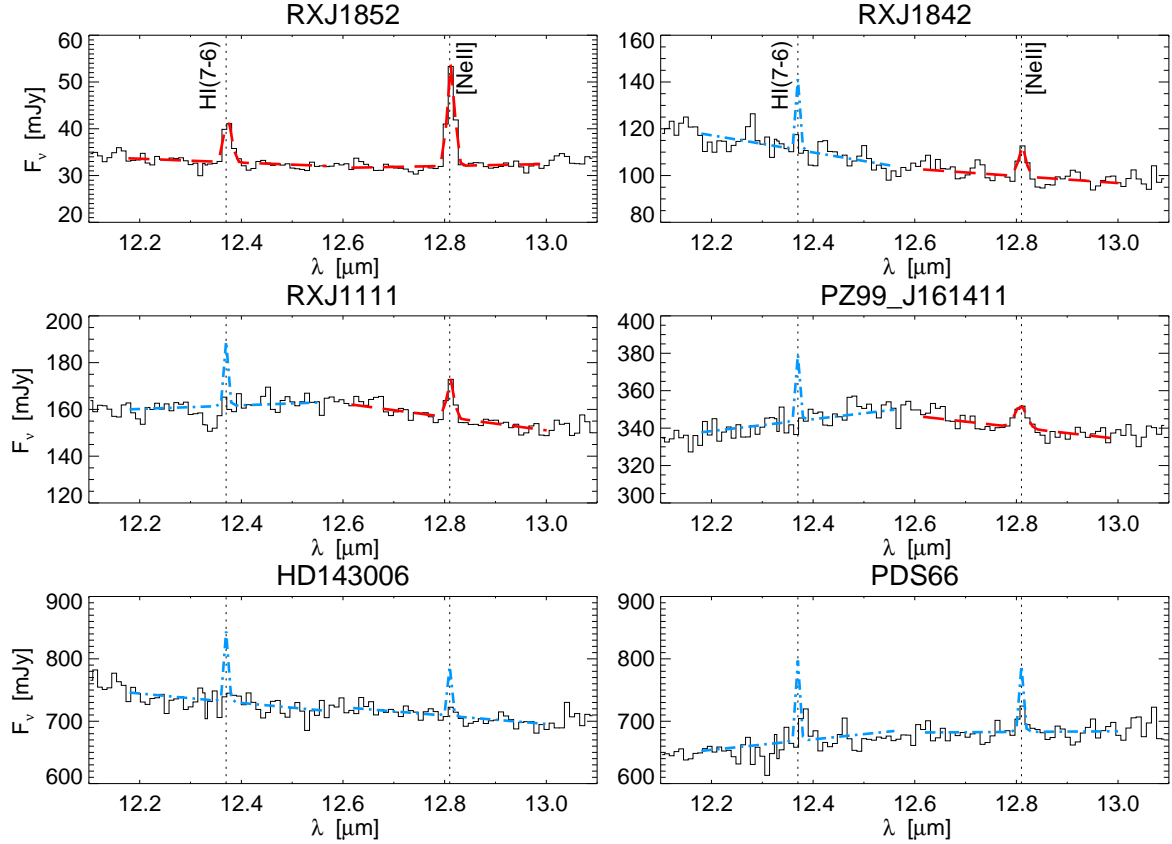}
\caption{Expanded view of the wavelength regions around the \hi(7--6) and \neii{} emission lines.
On top of the stellar and dust continuum we overplot the best Gaussian fits to the data (red dashed--lines) and the hypothetical 3$\sigma$ upper limits (light blue dot--dashed lines) reported in Table~\ref{T:Lines}. 
In the case of PDS~66, we might have detected \neii{} emission at a level of $\sim$2$\sigma$ (Flux$\,\sim \, 1.4 \times 10^{-14}{\rm erg}\, {\rm s}^{-1} {\rm cm}^{-2}$). However, due to the faintness of the emission we cannot confirm its presence in both nod positions and therefore prefer to report a 3$\sigma$ upper limit in Table~\ref{T:Lines}. 
\label{F:Lines}}
\end{figure*}

\begin{figure*}
\includegraphics[angle=90,scale=0.7]{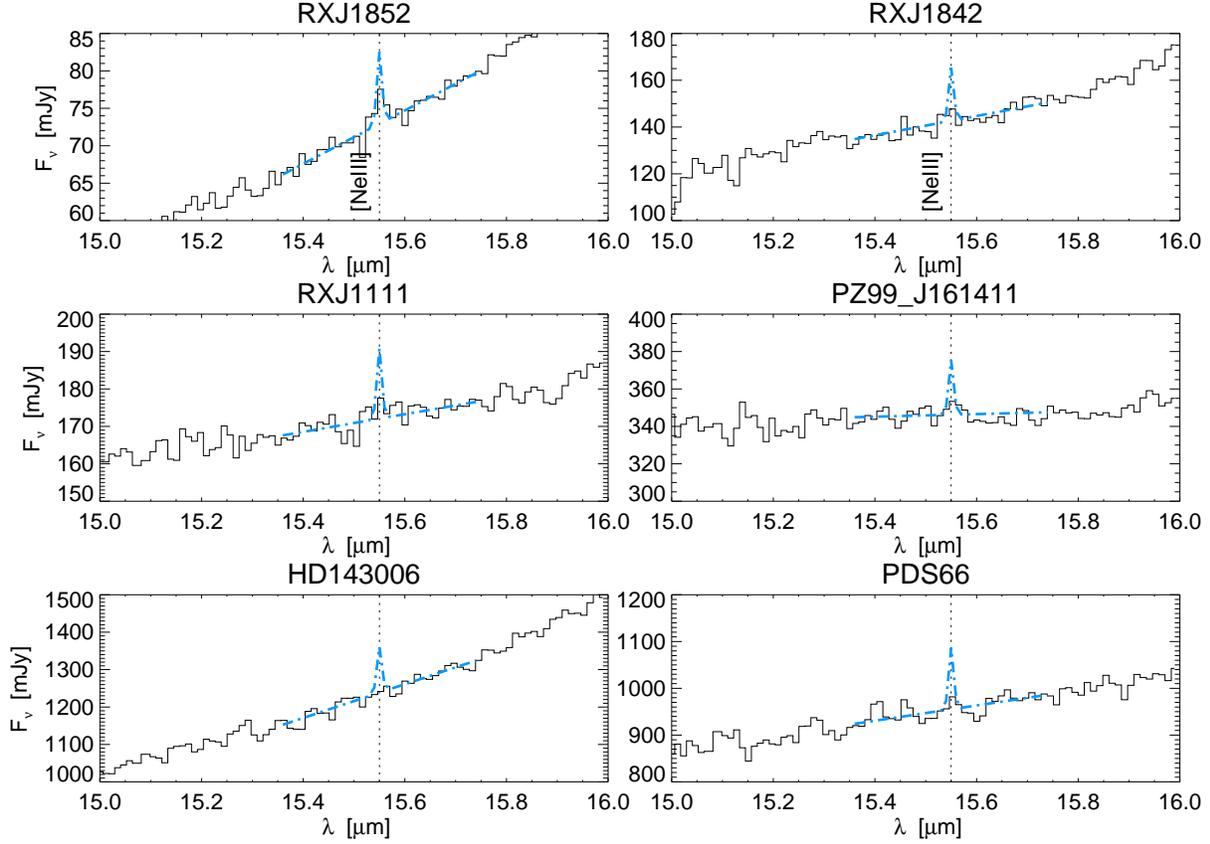}
\caption{Expanded view of the wavelength regions around the \neiii{} line at 15.55\,\micron .
On top of the stellar and dust continuum we overplot the hypothetical 3$\sigma$ upper limits 
(light blue dot--dashed lines) reported in Table~\ref{T:Lines}. \label{F:neiii}}
\end{figure*}

\begin{deluxetable}{l c c c}
\tabletypesize{\scriptsize}
\tablewidth{0pt}
\tablecaption{Line fluxes and upper limits (3$\sigma$) from the IRS/Spitzer spectra.\label{T:Lines}}
\tablehead{
\colhead{Source} & \colhead{Flux(\hi)} &\colhead{Flux(\neii)} &\colhead{Flux(\neiii)} \\
\colhead{} & \colhead{[10$^{-15}\,{\rm erg}\, {\rm s}^{-1} {\rm cm}^{-2}$]}& 
\colhead{[10$^{-15}\,{\rm erg}\, {\rm s}^{-1} {\rm cm}^{-2}$]} & 
\colhead{[10$^{-15}\,{\rm erg}\, {\rm s}^{-1} {\rm cm}^{-2}$]} 
}
\startdata
RX~J1111.7-7620 	&$<$5.6         & 5.1$\pm$1.2 & $<$3.0 \\
PDS~66	&$<$26.0        &$<$20.0	         & $<$21.1  \\ 
HD~143006            	&$<$22.6        &$<$15.0                 & $<$19.7 \\ 
$[$PZ99$]$~J161411.0-230536	&$<$7.2         & 6.2$\pm$1.8 & $<$4.7\\
RX~J1842.9-3532 	&$<$6.1         & 4.3$\pm$1.3 & $<$3.6\\
RX~J1852.3-3700 	&3.4$\pm$0.4    & 7.2$\pm$0.4            & $<$1.5\\
\enddata
\tablecomments{The wavelengths for the \hi (7-6), \neii , and \neiii{} lines are  12.37, 12.81, and 15.55\,\micron{} respectively. 
The 1$\sigma$ errors on the detections are calculated from the RMS of the observations minus model fit.}
\end{deluxetable}

\subsection{Fluxes of Infrared Gas Lines}\label{S:detections}
We report the detection of \neii{} emission at 12.81\,\micron{} in four out of
the six targets and the additional detection of \hi{}(7--6) at 12.37\,\micron{}
in RX~J1852.3-3700 (see Fig.~\ref{F:Lines}). For these detections, we verified
that the emission is centered at the expected rest wavelength, is spectrally
unresolved\footnote{The resolving power of the SH module is R$\sim$650
corresponding to $\sim$460\,km/s.}, is present in both nod1 and nod2 positions, and is
absent in the sky positions.  These checks guarantee that the emission originates
at the source location or in its proximity, within  2\farcs3/pixel or
$\sim$320\,AU at a distance of 140\,pc.

To estimate the fluxes of the detected lines, we fit the spectrum within
$\pm$0.25\,\micron{} of each line using a Levenberg--Marquardt  algorithm and
assuming a Gaussian for the line profile and a first--order polynomial for the
continuum.  The 1$\sigma$ errors on the line fluxes are evaluated from the RMS
dispersion of the pixels  in the spectrum minus the best fit model (see
Table~\ref{T:Lines}).  In the case of non--detections, we fit the same spectral
range with a first--order polynomial and we provide in Table~\ref{T:Lines} the  
3$\sigma$ upper limits to the flux from the RMS in the baseline subtracted
spectrum.  Figs.~\ref{F:Lines} and~\ref{F:neiii}  illustrate our best fits to the detected lines
(red dashed lines) and the hypothetical 3$\sigma$ upper limits (light blue dot--dashed lines).
In addition to the \hi{} and the \neii{} transitions, we provide upper limits for the \neiii{} line at 15.55\,\micron{}
for which we have predicted line emissions from the X--ray and EUV models.

\subsection{Mass Accretion Rates}\label{S:Mdot}
Observational evidence for magnetospheric
accretion in classical T Tauri stars (TTSs) is robust (e.g. \citealt{bouvier07}).
Evidence of this process are the UV/optical continuum excess, emitted
by the accretion shock on the stellar surface, and permitted emission lines originating
in the infalling magnetospheric gas (e.g. \citealt{muzerolle01,kurosawa06}).
The UV/optical continuum excess emission  
is the most common and direct measure of mass accretion rates 
($\dot{M_\star}$). However, in many instances, this excess is too weak to be measured, and other methods such as modeling Balmer emission profiles are necessary.
This is particularly true for objects with predominately low accretion rates, such as older 5--10\,Myr
TTSs (e.g. \citealt{muzerolle00,lawson04}) in the same age range as our sample.
%We estimated mass accretion rates ($\dot M$) for our sample by modeling
%Balmer emission profiles.  Previous work has shown that in all but
%the most active young T Tauri stars (TTSs), permitted line emission originates
%from magnetospheric accretion flows, in which material accreting
%through a circumstellar disk is funneled onto the stellar surface
%(e.g. \citealt{muzerolle01,bouvier06,kurosawa06}).
%Another by-product of this process, UV/optical continuum excess emitted
%by the accretion shock on the stellar surface, is the most common and
%direct method of determining $\dot M$.  However, in many instances,
%this excess is too weak to be measured, and other methods such as
%Balmer line modeling are necessary.  This has been particularly true for
%objects with predominately low accretion rates, such as older 5-10\,Myr
%TTSs (e.g. \citealt{muzerolle00,lawson04}) in the same age range
%as our sample.

\begin{figure*}
\plotone{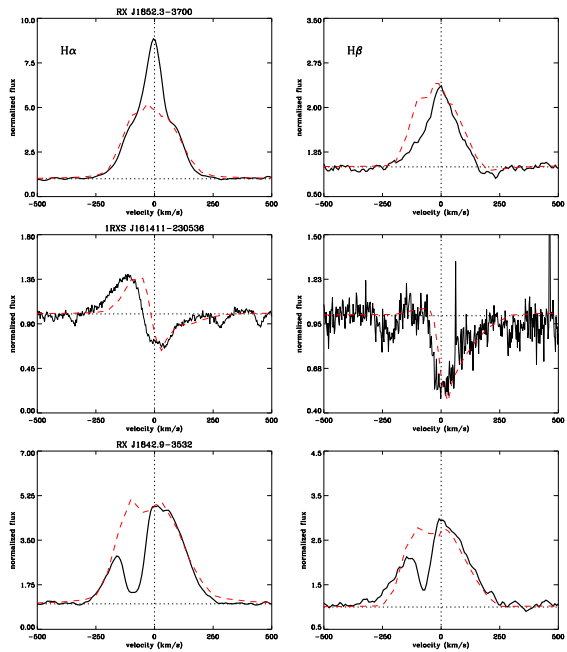}
\caption{Observed and model Balmer line profiles (solid and dashed lines,
respectively) for the sources RX~J1852.3-3700, $[$PZ99$]$~J161411.0-230536, and
RX~J1842.9-3532.  Model parameters are listed in Table~\ref{T:Accr}.
\label{profiles1}}
\end{figure*}

\begin{figure*}
\plotone{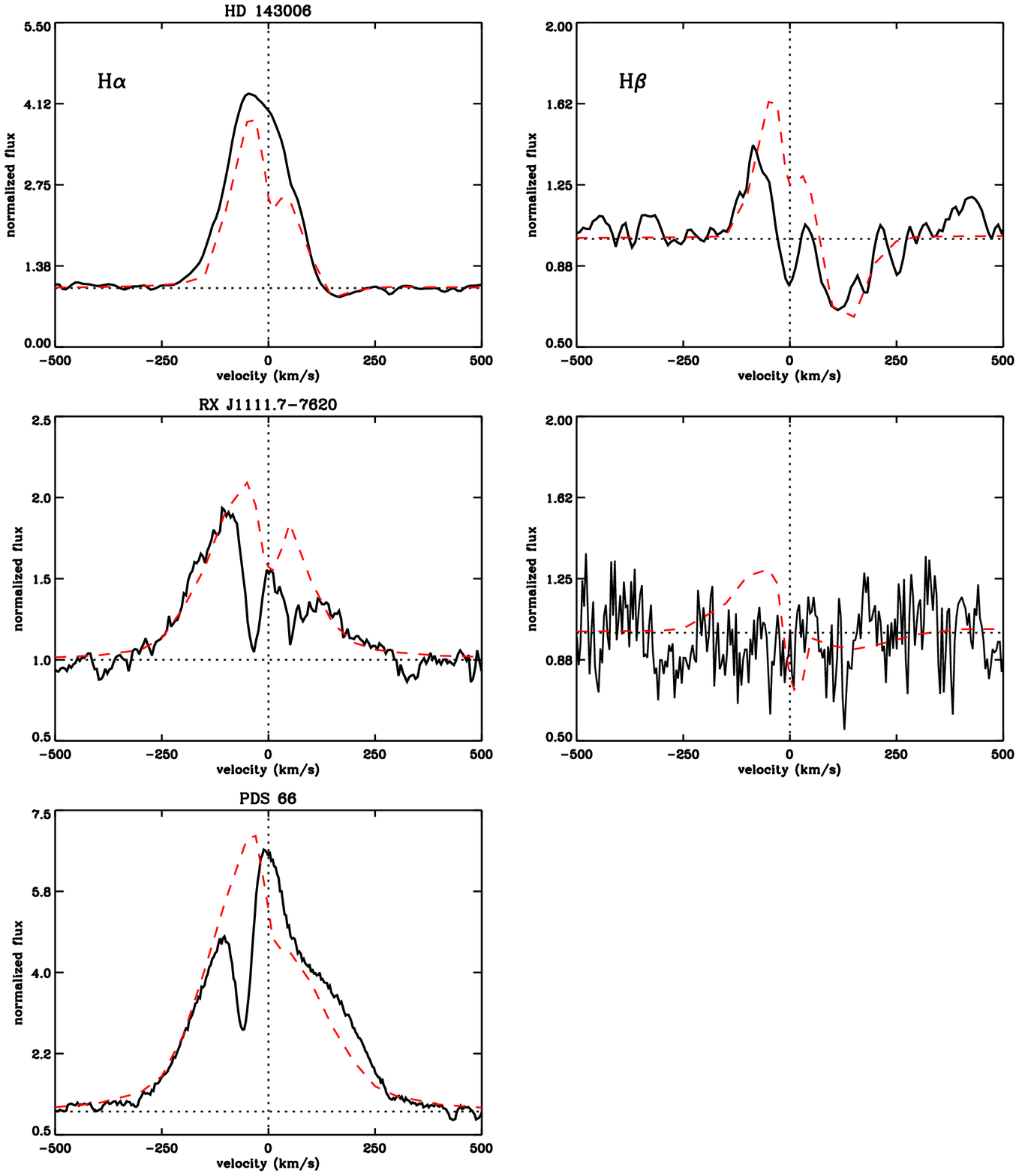}
\caption{Observed and model Balmer line profiles (solid and dashed lines,
respectively) for the sources HD~143006, RX~J1111.7-7620, and PDS~66.  
Model parameters are listed in Table~\ref{T:Accr}.
\label{profiles2}}
\end{figure*}

Many diagnostics in our spectra suggest low accretion rates in comparison
to values typical of younger TTSs such as in Taurus: the lack of optical
continuum veiling (continuum excess over photospheric flux $<$\,0.1), weak or absent mass loss signatures such as \oi{}
$\lambda$6300\,\AA{} emission, and the lack of broad emission components
in the Na~D doublet and \caii{} triplet lines.  The H$\alpha$ and H$\beta$
emission profiles, shown in Figures~\ref{profiles1} and \ref{profiles2}, are in most cases
qualitatively suggestive of weak accretion.  
One object, RX~J1842.9-3532, exhibits blueshifted absorption indicative of significant
mass loss, and two objects exhibit redshifted absorption components
in H$\alpha$ ([PZ99]~J161411.0-230536  and HD~143006).  Given the spectral types of these stars, the lack of optical
veiling  implies an upper limit of
$\dot{M_\star} \lesssim 10^{-8}$\,M$_{\odot}$\,yr$^{-1}$ (see \citealt{calvet04}).

We followed the procedures outlined in detail
in \citet{muzerolle01} to model the Balmer line profiles and thus estimate mass accretion rates for our targets.
Although the paper by \citet{muzerolle01} focused on low--mass stars with late K 
and M spectral types, accretion diagnostics such as UV excess and Pa$\beta$ 
(\citealt{calvet04}, \citealt{natta06a}) 
indicate that magnetospheric model assumptions can be extended to stars 
with early K and G spectral types like our targets.
We adopted stellar parameters based on
the empirically--derived quantities in Table~\ref{T:Stars}.  
Gas temperatures were set roughly to 10,000\,K following 
the constraints derived in \citet{muzerolle01}.
Note that for the density regime appropriate for these particular objects,
the gas is already nearly fully ionized and larger gas temperatures will not
result in significantly different line emission.  The magnetosphere size/width
was set to a fiducial value (between 2.2--3\,$R_\star$), lacking any observable constraints.  
This is the largest source of uncertainty in constraining the accretion rate;
however, a plausible range of values results in no more than a factor of 3--5
range in accretion rates that can reproduce the observed line profiles.
We further included rotation, using the treatment of \citet{muzerolle01}, since rotation rates
typical of our objects can have an observable effect on the line profiles near the line center.
The adopted stellar equatorial velocities, based on the observed $v~sin(i)$ values, are given in Table~\ref{T:Accr}.
Finally, the model inclination angle and mass accretion rate are varied
to reproduce the wings, width, and peak of the observed lines.  The best matches
are shown in Figs.~\ref{profiles1} and \ref{profiles2} and the inferred inclination angles ($i$) and mass accretion 
rates ($\dot{M_\star}$) are summarized in Table~\ref{T:Accr}. Note that $\dot{M_\star}$
$\la5\times10^{-11}$\,M$_{\odot}$\,yr$^{-1}$ produce negligible H$\alpha$ emission for our sources
and thus represent lower limits for their accretion rates. Given the magnetospheric model parameters
in Table~\ref{T:Accr}, we also compute the predicted flux in the \hi{}(7--6) transition (last column of the table).

The models reasonably account for the observed profiles,
with a few exceptions.  The H$\alpha$ emission line of
RX~J1852.3-3700 exhibits a narrow, symmetric core that the models
cannot reproduce. 
%this feature is strangely absent from the H$\beta$ line.
It is possible that there is a chromospheric emission component superposed
on top of the broader accretion component.  For this to occur,
a significant amount of the stellar surface must not be covered by
the accretion flow (which would be consistent with the more pole-on
orientation of the model).  The models also cannot reproduce the blueshifted
absorption components seen in both H$\alpha$ and H$\beta$ profiles of
RX~J1842.9-3532.  A treatment of the wind is necessary to account for
such features, which is beyond the scope of this work.  However, we note that
the models still account for the overall emission profile, suggesting
that the wind produces negligible emission.

The derived mass accretion rates  are all lower than the average value for
1 Myr--old T~Tauri stars \citep{gullbring98,calvet04}
by factors larger than 10.  Given the older ages of the stars
in our sample, this may be consistent with a general decline in accretion rate
with time as predicted by models of viscous disk evolution
(e.g. \citealt{hartmann98,muzerolle00}). 
The predicted  \hi{}(7--6) emission for RX~J1852.3-3700  contributes to at most $\sim$30\% of the observed value\footnote{This estimate includes a factor of 2 uncertainty in the predicted \hi{} flux, see Note to Table~\ref{T:Accr}}, while predicted fluxes from the other sources are negligible or  fall well below the 3$\sigma$ upper limits we report in Table~\ref{T:Lines}.
Thus, magnetospheric accretion models argue against the  \hi{}(7--6) transition originating in accretion flows.  Accretion shock regions, the stellar chromosphere, and the hot disk surface are other possible sources of emission. We discuss the contribution from the disk in the next Section and note that the possible chromospheric emission component in the  H$\alpha$ profile of RX~J1852.3-3700 could also contribute to the \hi{}(7--6) line.

\begin{deluxetable}{lcccccc}
\tabletypesize{\scriptsize}
\tablewidth{0pt}
\tablecaption{Magnetospheric accretion model parameters and resulting mass accretion rates ($\dot{M_\star}$).  \label{T:Accr}}
\tablehead{
\colhead{Source} & \colhead{$R_*$} &
%\colhead{$T_{\rm{eff}}$} & 
\colhead{$V_{\rm eq}$} &
\colhead{$T_{\rm{max}}$} & \colhead{$i$} &
\colhead{log($\dot{M_\star}$)} & \colhead{Predicted Flux(\hi)} \\
\colhead{} & \colhead{[$R_{\odot}$]} & 
%\colhead{[K]} & 
\colhead{[km\,s$^{-1}$]} &
\colhead{[K]} & \colhead{[\degr]} & 
\colhead{[M$_{\odot}$\,yr$^{-1}$]} & 
\colhead{[10$^{-15}\,{\rm erg}\, {\rm s}^{-1} {\rm cm}^{-2}$]} 
}
\startdata
%RX~J1111.7-7620          & 2.0 & 4600 & 25 & 12000 & 75 & -9.3 & 0\\
%PDS~66                         & {\bf ERIC!} &     &      &            &      &       &    \\ 
%HD~143006                   & 1.5 & 5900 & 10 & 12000 & 10 &  -9.3&  1.3  \\
%$[$PZ99$]$~J161411.0-230536& 2.5 & 5000 & 45 & 12000 & 65 & -9.5 & 0 \\
%RX~J1842.9-3532         & 1.0 & 4800 & 25 & 10000 & 45 & -9.0 & 0.48 \\
%RX~J1852.3-3700         & 1.0 & 4800 & 25 & 12000 & 35 & -9.3 & 0.56\\
%=======================================================
RX~J1111.7-7620          & 2.0  & 25 & 12000 & 75 & -9.3 & 0\\
PDS~66                   & 1.0  &  0 & 9000  & 45 & -8.3 & 2.4    \\ 
HD~143006                   & 1.5  & 10 & 12000 & 10 &  -9.3&  1.3  \\
$[$PZ99$]$~J161411.0-230536& 2.5  & 45 & 12000 & 65 & -9.5 & 0 \\
RX~J1842.9-3532         & 1.0 & 25 & 10000 & 45 & -9.0 & 0.48 \\
RX~J1852.3-3700         & 1.0 & 25 & 12000 & 35 & -9.3 & 0.56\\
\enddata
\tablecomments{$T_{\rm{max}}$ is the maximum value of the adopted temperature
distribution of the accretion column.  All models are calculated for a solar mass star and for
outer magnetospheric radii between 2.2 and 3\,$R_*$.
The last column gives  the predicted flux in the \hi{}(7--6) transition at 12.37\,\micron .
Uncertainties in the mass accretion rates are no more than a factor of 3--5. 
By varying $\dot{M_\star}$ by a factor of 5 the models give a range of \hi{} fluxes within a factor of $\sim$2.
}
\end{deluxetable}

\section{Disk Atmosphere and \neii{} Emission}\label{S:origin}
Two-thirds of our optically thick dust disks present a \neii{} emission line at 12.81\,\micron{}.
Where does the \neii{} emission originate? 
Recently \citet{glass07} showed that stellar X-rays can partially ionize the gas in the disk atmosphere
and produce detectable \neii{} lines. 
Alternatively, Hollenbach \& Gorti in prep. suggest that extreme ultraviolet (EUV, $h\nu > 13.6$\,eV) photons from the central star
ionize the upper layer of circumstellar disks and create a kind of coronal \hii{} region producing detectable \neii{} emission.
 Although the two models rely on different ionization mechanisms, they both
predict that \neii{} emission originates from gas in a hot surface layer of circumstellar disks\footnote{The coronal \hii{} region lies on top of the slightly lower X--ray heated region, which is only partially ionized ($\sim$0.1 to 1\%).}.
We can thus expect to find some correlations between the observed \neii{} fluxes and the star/disk properties such as infrared excess emission, stellar UV or X--ray flux, and mass accretion rates.

\begin{figure}
\includegraphics[angle=0,scale=0.45]{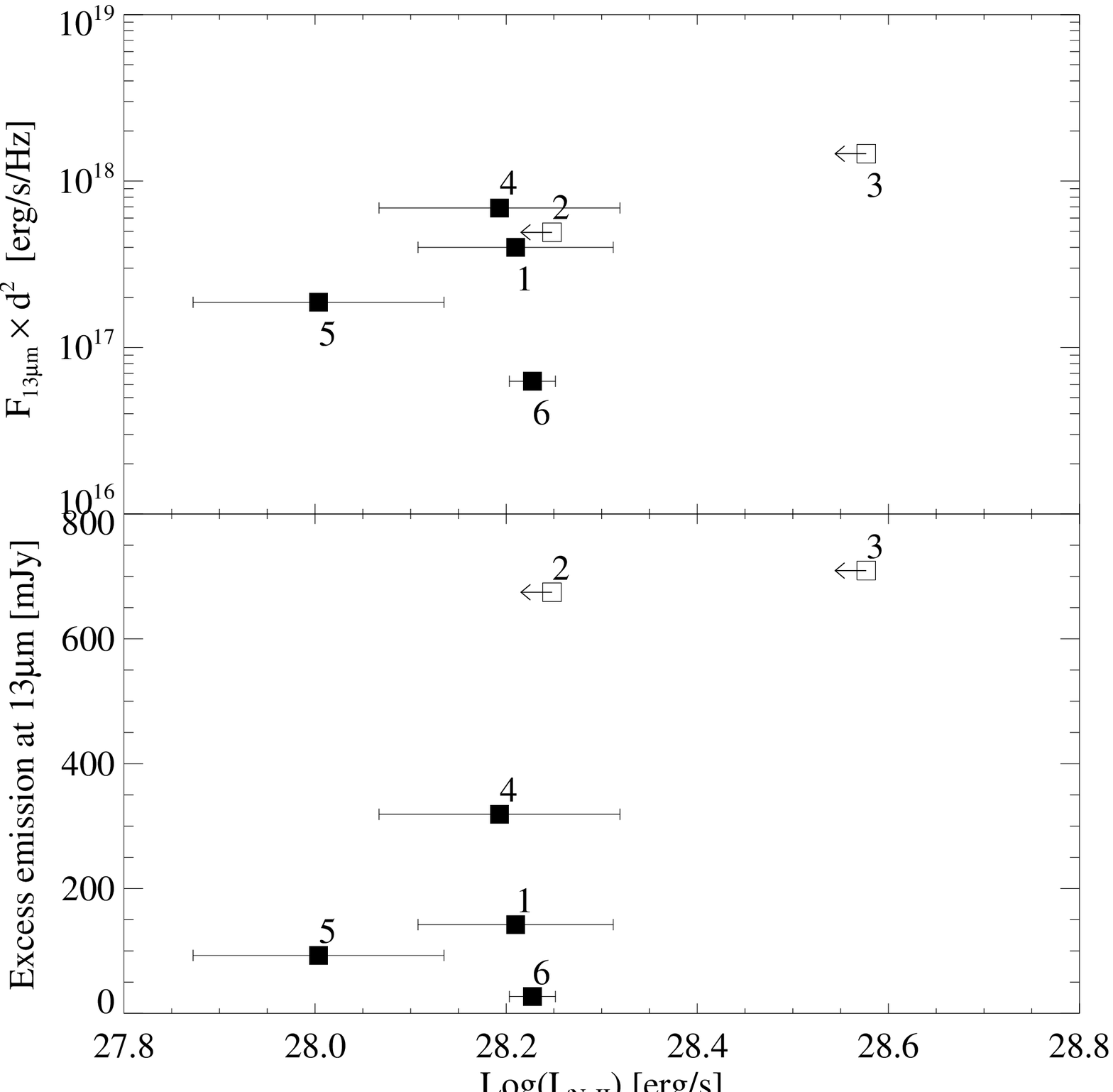}
%\plotone{correl_disk.ps}
\caption{Possible correlations of \neii{} line luminosities with disk properties.
\neii{} detections are filled squares and right--most \neii{} limits are open squares with left pointing arrows.
The numbers correspond to the ID numbers of our targets as given in Table~\ref{T:Stars}.
Upper panel: Dereddened flux between 12.85 and 13.15\,\micron{} (as estimated from the low--resolution spectra,
\citealt{bouwman07}) times distance squared versus the \neii{} line luminosity. To deredden the fluxes we used  the visual extinctions in Table~\ref{T:Stars} and the mean extinction law from \citet{mathis90}.
Lower panel: Excess emission at 13\,\micron{} versus the \neii{} line
luminosity. The excess emission is computed from the difference between the observed IRS low--resolution flux between 12.85 and 13.15\,\micron{}  and the best fit Kurucz model atmospheres (see \citealt{meyer06} for details on the procedure) in the same wavelength range. Excess emission has been dereddened as explained above. In both plots the errorbars in the y--axis are few percent of the plotted values and thus smaller than the used symbols. 
Low infrared excess emission makes it more favorable to detect \neii{} emission lines.\label{F:disk}}
\end{figure}

First we explore any correlation between the continuum emission in the vicinity of the \neii{} line
and the \neii{} line luminosities (see Fig.~\ref {F:disk}).  Although there is no obvious trend between the plotted quantities, it is interesting that the four detections cluster in a narrow range of \neii{} line 
luminosities (differences less than a factor of 2) and below 400\,mJy of excess emission at 13\,\micron .
The weaker mid--infrared continuum level (caused by an inner hole and/or grain growth)  improves the line to continuum ratio in these relatively low spectral resolution {\it Spitzer} observations and thus 
makes it more favorable to detect \neii{} emission lines. 
%Note that the strongest
%detection is from RX~J1852.3-3700 which has the lowest excess emission at 13\,\micron , whereas the two \neii{} upper limits are from the sources with the largest excess emission. 

\begin{figure}
\includegraphics[angle=0, scale=0.45]{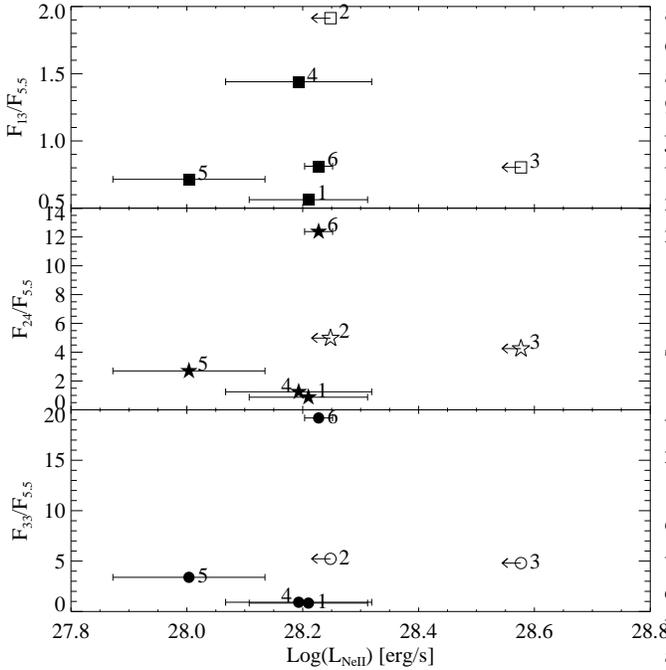}
%\plotone{correl_disk.ps}
\caption{Possible correlations of \neii{} line luminosities with the disk flaring. 
\neii{} detections are filled symbols and right--most \neii{} limits are open symbols with left pointing arrows.
The disk flaring is represented by the flux ratio at 13 (upper panel), 24 (middle panel), and 33\,\micron{} (lower panel) over the reference flux at 5.5\,\micron , that is the shortest wavelength covered by the IRS low--resolution spectra. More flaring is indicated by higher ratios in the figure.
There is no obvious correlation between disk flaring and \neii{} emission. \label{F:flaring}}
\end{figure}

We also search for correlations with the disk structure, in particular with the disk flaring.
The continuum flux emitted from the surface layer of the disk is proportional to the angle at which  stellar radiation impinges onto the disk.
This so--called grazing angle becomes proportional to the disk flaring at radial distances from solar--type stars $\ga$0.4\,AU \citep{cg97}. Such distances are probed by dust emission at mid--infrared wavelengths. Thus, we can use the ratio of fluxes at two different mid--infrared  wavelengths to trace changes in the disk flaring.  We chose as reference flux that at 5.5\,\micron{}, which is the shortest wavelength covered by the IRS low--resolution modules, and computed the flux ratios at 13, 24 and 33\,\micron{} (see Fig.~\ref{F:flaring}). Larger flaring is indicated
 by higher ratios of long wavelength continuum to short wavelength continuum
 flux from the dust disk. Source RX~J1852.3-3700 is peculiar in having small excess at 13\,\micron{} and large excesses at 24 and 33\,\micron . The SED of this object \citep{silverstone06} is reminiscent of transition systems such as GM~Aur and DM~Tau \citep{calvet05}.
We do not see any obvious correlation between disk flaring and \neii{} line luminosities. 

\begin{figure*}
\includegraphics[angle=0,scale=0.7]{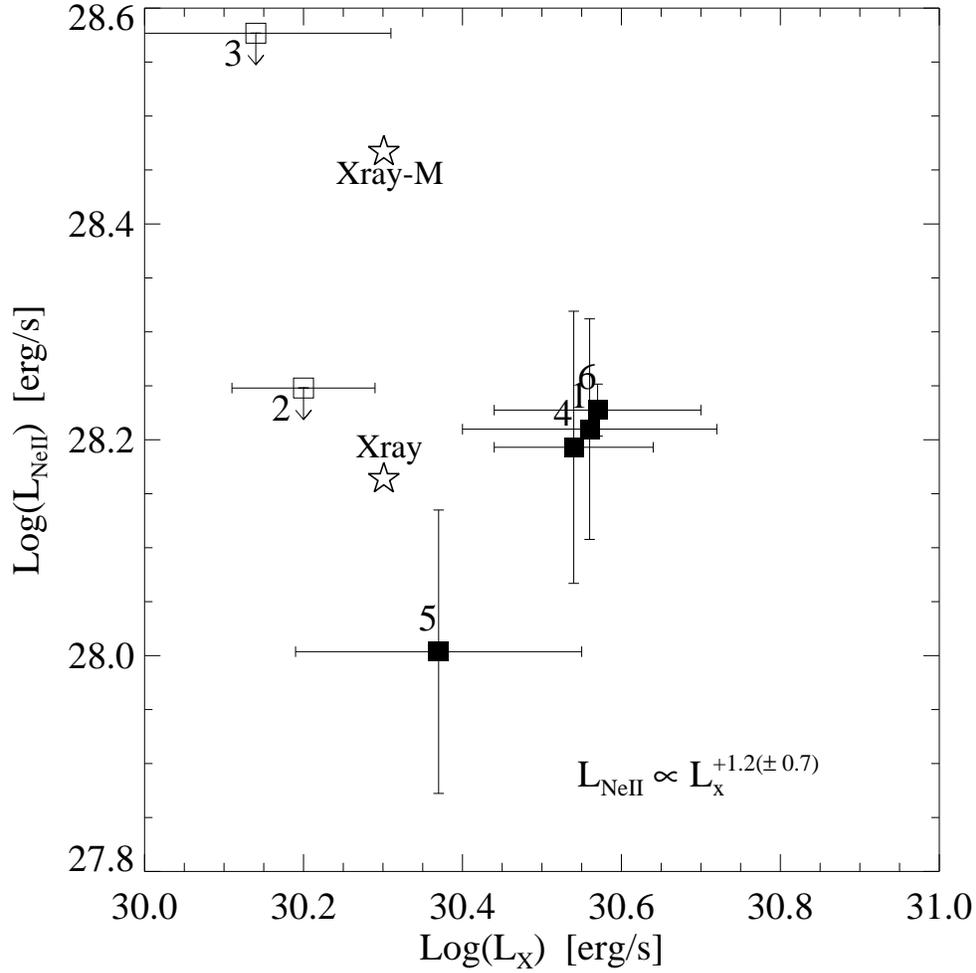}
\caption{Line luminosities (filled symbols) and upper limits (open symbols with downward arrows) 
for the \neii{} transition versus the star X-ray
luminosity as given in Table~\ref{T:Stars}. The two open stars represent the two extreme thermal models from
\citet{glass07}: the upper model is when mechanical heating (accretion) dominates (Xray-M), the lower model is when X--ray heating dominates (Xray). The ordinary least--squares fit to the four detections suggests a positive correlation between $L_{\rm [Ne~{\sc II}]}$ and $L_{\rm X}$. \label{F:Xray}}
\end{figure*}

\begin{figure*}
\includegraphics[angle=0,scale=0.7]{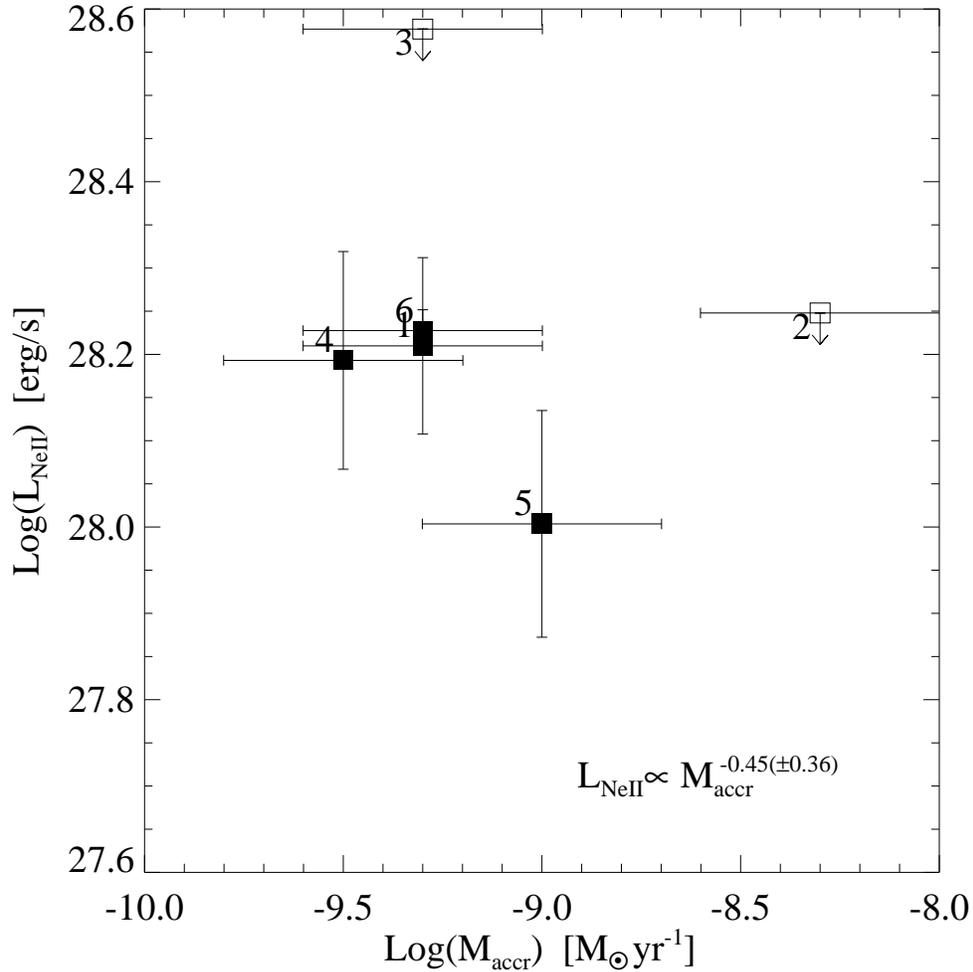}
\caption{Line luminosities (filled symbols) and upper limits (open symbols with downward arrows) 
for the \neii{} transition versus the stellar accretion rates
given in Table~\ref{T:Accr}.  We overplot a factor of 4 uncertainty in the mass accretion rates,  representative of the uncertainties we estimate from the models (see Sect.~\ref{S:Mdot}). The ordinary least--squares fit to the four detections suggests an anti--correlation between $L_{\rm [Ne~{\sc II}]}$ and $\dot{M_\star}$. \label{F:Mdot}}
\end{figure*}

Interestingly, we find trends with the stellar X--ray luminosities (Fig.~\ref{F:Xray}) and
with the mass accretion rates from the Balmer profiles (Fig.~\ref{F:Mdot}).
Stellar X--ray luminosities are computed from the ROSAT All--Sky Survey count rates and hardness ratios
(see Table~\ref{T:Stars}). We did not attempt to correct for the interstellar extinction because the X--ray spectrum of our targets is not well--known from the ROSAT observations alone. 
Nevertheless, the largest difference in visual extinction (0.7\,mag between HD~143006 and  RX~J1852.3-3700) translates into less than 0.1\,dex difference in log($L_{\rm X}$), well within the estimated errors and variations due to stellar activity. This indicates that any observed trend in our sample can be only marginally influenced by the correction in interstellar extinction. The best (minimum chi--square) fit to the \neii{} detections in  Fig.~\ref{F:Xray} gives the following correlation between the \neii{} and the X--ray luminosity:
\begin{displaymath}
L_{\rm [Ne~{\sc II}]} \propto   L_{\rm X}^{+1.2(\pm 0.7)}
\end{displaymath}
The slope of the relation is not constrained  but it is positive within 1.5$\sigma$. To further evaluate the likelihood of  a positive correlation, we generate 100 normally--distributed points around the four \neii{} detections and fit the data using six different linear regression methods described in \citet{isobe90}. The analytical slopes as well the average slopes of 100 random resampling of the data using bootstrap and  jackknife techniques \citep{babu92,feigelson92} are all positive.  The Pearson correlation coefficient is 0.24 meaning that we can be confident of a positive correlation between  $L_{\rm [Ne~{\sc II}]}$ and $L_{\rm X}$ at a 95\% confidence level.
We perform a similar analysis to test the apparent anti--correlation between the \neii{} luminosity and the mass accretion rate in Fig.~\ref{F:Mdot}. The ordinary least--squares fit to the four \neii{} detections results in:
\begin{displaymath}
L_{\rm [Ne~{\sc II}]} \propto   \dot{M_\star}^{-0.45(\pm 0.36)}
\end{displaymath}
The slope of the relation is negative only within 1$\sigma$.
The Monte Carlo approach in combination with the six regression methods also indicates a negative correlation at a 95\% confidence level.

In summary, we have evidence for a positive correlation between $L_{\rm [Ne~{\sc II}]}$ and $L_{\rm X}$ and a negative correlation between $L_{\rm [Ne~{\sc II}]}$ and $\dot{M_\star}$. However we cannot quantify the slopes of these relations because of the small number of detections and of the large errorbars in the measured and estimated quantities. It is also important to point out that source 5 (RX~J1842.9-3532) has a large weight in determining these trends being the target with the most different X--ray luminosity and mass accretion rate. 
%We were informed of an additional \neii{} detection from the  TW~Hya system, which has also different $L_{\rm X}$ and $\dot{M_\star}$ from our sample,  following the correlations identified here (Joan Najita private communication). 
Note that if the 2$\sigma$ detection from PDS~66 is real it would confirm the positive trend between \neii{} luminosity and X--ray luminosity.
A larger sample of sources spanning a wider range in X--ray luminosities and mass accretion rates is necessary to  constrain the slopes of the $L_{\rm [Ne~{\sc II}]}$--$L_{\rm X}$ and $L_{\rm [Ne~{\sc II}]}$--$\dot{M_\star}$ relations.
In the subsections below we discuss predictions from the X--ray and EUV models and compare them to 
the measured  \neii{} fluxes and to the correlations presented in this Section.

\subsection{Plausible Disk Origin for the Ne$^+$ emission}
According to the model proposed by \citet{glass07}, 
detectable \neii{} emission can be produced by the atmosphere of a disk 
both ionized and heated by stellar X--rays \citep{glass04}.    
Ne ions, primarily Ne$^+$ and Ne$^{2+}$, are 
generated through X--ray ionization and destroyed by charge exchange 
with atomic hydrogen and radiative recombination.
Because the X--rays emitted by young stars have a characteristic
energy similar to the K--edges of Ne and Ne$^+$ at $\sim 0.9$\,keV,
they are efficient in producing Ne ions.  The warm disk surface 
region ($T \sim 4000$\,K) generated by X--ray irradiation extends 
out to large radii ($\sim 20$\,AU).  As a result, fine structure 
transitions of Ne$^+$ and Ne$^{2+}$, which have excitation temperatures 
of $\sim 1000$\,K, can be produced over a vertical column of warm gas of 
$10^{19}-10^{20}$\,cm$^{-2}$ and over a large range of disk radii. 
These circumstances lead to a significant flux of \neii{} at 
12.81$\mu$m and \neiii{} at 15.55$\mu$m.

For a typical T Tauri disk \citep{dalessio99} located at 140\,pc, \citet{glass07} predict that a 
stellar X--ray luminosity of log$(L_{\rm X}) = 30.30$\,erg\,s$^{-1}$ 
generates a \neii{} flux of $6.22\times 10^{-15}$\,erg\,s$^{-1}$\,cm$^{-2}$  
when accretion related processes 
are unimportant in heating the disk surface.  
The \neiii{} flux is estimated to be approximately 10 times lower.
The X--ray luminosity assumed in the Glassgold et al. model
is close to that of RX~J1842.9-3532 (see Table~\ref{T:Stars} and Fig.~\ref{F:Xray}) and the 
predicted \neii{} flux is similar (within a factor of 1.5) to the value we report in Table~\ref{T:Lines}.
Unfortunately, the upper limits on the \neiii{} lines (see Table~\ref{T:Lines}) are not 
stringent enough to test the model predictions.

In the model proposed by Hollenbach \& Gorti (2007, in preparation),
EUV photons from the stellar chromosphere \citep{alexander05} and/or from accretion 
\citep{matsuyama03,herczeg07} create an \hii --region like ionized layer
on the surface of young circumstellar disks. The EUV photons incident
on the disk are mostly absorbed near the so--called ''gravitational radius''
($\sim 10$\,AU for a 1\,M$_{\odot}$ star) where the local thermal speed of
the $10^4$\,K ionized hydrogen nuclei is equal to the escape speed from the
gravitational potential (e.g. \citealt{holl94}). The numerical
model for the EUV--heated disk surfaces includes a calculation of
the diffuse field due to hydrogen and helium recombinations,
and ionized gas chemistry comprising photoionizations, recombinations,
and charge exchange reactions (Hollenbach \& Gorti 2007, in prep.).
In the case of a soft EUV spectrum (e.g. a blackbody at $4\times 10^4$\,K),
photon luminosities of $\sim 10^{41}$ erg s$^{-1}$  produce a \neii{} line luminosity 
of $\sim 10^{-6}$\,L$_{\odot}$,  which corresponds to a flux of $2\times10^{-15}$ erg s$^{-1}$
cm$^{-2}$ from a source at 140\,pc.  Harder EUV spectra can produce
more doubly ionized Ne in the disk atmosphere which may result in stronger \neiii{} lines
in comparison to the case of soft EUV spectra (Hollenbach \& Gorti 2007, in prep.).

With the assumption that Ne atoms are ionized only by stellar EUV photons and that the sources have a soft EUV spectrum, one can use \neii{} lines as an indirect tool to estimate  EUV fluxes\footnote{In the EUV model the \neii{} luminosity is directly proportional to the EUV luminosity of sources with soft EUV soft spectra}, which are unconstrained for the majority of the stars.
The \neii{} fluxes we report in Table~\ref{T:Lines} convert to EUV fluxes between 2.6$\times 10^{41}$ (for RX~J1842.9-3532)  and 4.4$\times 10^{41}$ (for RX~J1852.3-3700) photons s$^{-1}$.  Such ionizing rates seem plausible for $\sim$5\,Myr old TTSs like our targets. For comparison  
\citet{herczeg07} calculate $\sim 10^{41}$ ionizing photons s$^{-1}$ for the $\sim$10\,Myr old TW~Hya system while
\citet{alexander05} estimate  a wide range of ionizing fluxes $\sim\,10^{41}-10^{44}$ photon s$^{-1}$  for a sample of five  classical T~Tauri stars. The estimates from \citet{alexander05} are based on modeling UV emission lines such as C~{\sc iv} and have only an order of magnitude accuracy due to  model uncertainties and more significanly to uncertainties in the reddening. 
Nevertheless, if X--rays contribute to ionize Ne atoms, as proposed by \citet{glass07}, our EUV estimates from \neii{} lines can be only taken as upper limits. 
%Similarly to the X--ray model, ionization of Ne atoms from EUV photons results in \neiii{} lines about 10 times weaker than  \neii{} emission lines and thus well below our 3$\sigma$ upper limits (see Table~\ref{T:Lines}). 

In summary, both the X--ray and the EUV models can reproduce the observed \neii{} line luminosities
with star/disk properties that are plausible for our targets. Can the same models explain the \hi (7-6) flux from 
RX~J1852.3-3700?
Hollenbach \& Gorti  estimate that L$_{\rm H~{\sc I}}\sim 10^{-3}$L$_{\rm [Ne~{\sc II}]}$ if 
all the \hi{}(7--6) emission originates from the ionized disk surface. They show that if X--rays dominate the \neii{} emission,
then the \hi{} luminosity should be even lower because the gas is mostly neutral. 
This demonstrates that neither the EUV nor the X--ray models can account for the observed \hi{} flux, which is only a factor of $\sim$2 lower than the \neii{} flux.
We conclude that the \hi{}(7--6) transition is not associated with either the
X--ray excited gas or the EUV excited gas in the disk. Because magnetospheric accretion flows can
account for at most $\sim$30\% of the observed \hi{}(7--6) flux  (Sect.~\ref{S:Mdot}), accretion shocks and/or the stellar chromosphere are left as possible major sources of emission.

\subsection{Predicted Correlations and Observed Trends}
In this subsection we discuss whether the trends identified in Sect.~\ref{S:origin} are consistent with predictions from the X--ray and EUV models.

First we consider the correlation between the \neii{} luminosity and the X--ray luminosity.
The X--ray model predicts that the ionization fraction of Ne atoms is proportional to the square root of the stellar X--ray luminosity:  $x(Ne^+)/x(Ne) \propto L_{\rm X}^{1/2} $ (eqs.~2.9 and 2.10 from \citealt{glass07}).
The integrated flux of the \neii{} line further depends 
on the excitation of the line, which is a function of the disk temperature and electron density. Further modeling by Meijerink, Glassgold \& Najita in prep. indicate that there is a 
 close to linear relation between $L_{\rm [Ne~{\sc II}]}$ and $L_{\rm X}$, which is consistent with the positive correlation suggested by our data.
This trend could be also consistent with the EUV model. In the case of a soft EUV spectrum, most of the Ne in the disk surface is Ne$^+$ and the \neii{} luminosity is predicted to be directly proportional to $L_{\rm EUV}$ (Hollenbach \& Gorti 2007 in prep.).
What is the relation between $L_{\rm EUV}$ and $L_{\rm X}$?
Emission lines such as \civ{} and \ovi{} tracing the EUV emission are 
found to scale with the X--ray flux with power law indices close to 0.5 in $\sim$Gyr old sun--like stars \citep{ayres97, guinan03}. If these relations hold for younger sun--like stars then the EUV model would also predict a positive correlation between $L_{\rm [Ne~{\sc II}]}$ and $L_{\rm X}$.
%============

\citet{glass07} also explored the effect of accretion heating on the \neii{} flux.
Accretion at a level of 10$^{-8}$\,M$_\sun$\,yr$^{-1}$, which is typical for a
classical TTS, results in a factor of 2 higher \neii{} flux than that produced by a  
non--accreting TTS surrounded by the same circumstellar disk (see also Fig.~\ref{F:Xray}).
This suggests a mild positive correlation between  $L_{\rm [Ne~{\sc II}]}$ and $\dot{M_\star}$, which is
different from the one we see in our data (Sect.~\ref{S:origin} and Fig.~\ref{F:Mdot}). 
However a detailed comparison with the models needs 
to take into account other factors  in addition to stellar accretion such as disk flaring and inclination.
In the case of the EUV model, stellar winds greater than $\sim 10^{-10}$\,M$_\sun$\,yr$^{-1}$  could
drastically reduce the EUV photons reaching the disk (Hollenbach \& Gorti in prep.). Given that the  
ratio of mass outflow rate to the mass accretion rate is close to $\sim 0.1$ for most classical TTSs with a large uncertainty \citep{shu00,white04}, these wind rates correspond to $\dot{M_\star} \ga 10^{-9}$\,M$_\sun$\,yr$^{-1}$.
Thus, the EUV model predicts significantly lower \neii{} luminosities for classical TTSs simply due to their higher accretion (and thus wind) rates.

Finally, disk flaring is also expected to play a major role in the X--ray model in that more flaring facilitates the penetration of stellar X--ray in the disk atmosphere \citep{glass07}. On the contrary, the \neii{} emission produced by EUV photons is expected to be independent of the disk flaring (Hollenbach \& Gorti in prep.). 
Quantitative predictions from both models are necessary before attempting any comparison with the empirically derived signature of disk flaring presented in Fig.~\ref{F:flaring}.

\section{Discussion}\label{S:discussion}
We have shown in the previous Sections that both the X-ray and EUV models can account for the
observed \neii{} fluxes suggesting that the emission originates in the hot disk atmosphere.
Other sources of \neii{} emission that we have not discussed so far might be a jet or an outflow. In this case, the
jet/outflow needs to provide enough energy to ionize Ne atoms, i.e. about twice the energy to ionize
atoms like S and N whose forbidden lines are sometimes detected in classical TTSs \citep{hartigan95}.
Our sources however have very weak or absent mass loss signatures and accretion rates at least a factor
of 10 lower than those of typical TTSs (see Sect.~\ref{S:Mdot}). Thus, we regard the possibility of a jet/outflow dominating the
\neii{} emission as very unlikely for our targets.  
%A comparison of our \neii{} fluxes to those from younger systems with high mass accretion rates and known jets/outflows might help  of their \neii{} emission could come from a jet/outflow.

It is possible to confirm that  \neii{} emission originates from the disk atmosphere by spectrally resolving the
detected lines. High--resolution spectroscopy could also enable the detection of the weaker \neiii{} line
that can be used  to further test the model predictions.
 Emission arising from disk radii out to $\sim 10$\,AU around a solar mass star would 
produce a line with $\sim 10$\,km/s width,
about 40 times smaller than the spectral resolution of the 
{\it Spitzer} IRS.  Current ground--based spectrographs with spectral 
resolutions $R\ga 30,000$ at 12\,\micron{} such as VISIR/VLT and TEXES/Gemini 
should be able to resolve such emission  lines. In addition similar observations of the \hi (7-6) line at 12.37\,\micron{} could confirm that the \hi{} line originates in a different region from the \neii{} line.

However high--resolution  spectroscopy alone is not sufficient to identify whether X--rays or EUV photons dominate the ionization of Ne atoms. In fact both the X--ray and EUV model predict a similar 
extension of the \neii{} emitting region.
Constraints for the EUV model can come from independent estimates of the stellar EUV flux.
Because EUV emission is produced by gas at temperatures between $10^5-10^6$\,K, an estimate of the emitting plasma
can be achieved by using gas lines such as the \ovi{} tracing gas at intermediate temperatures. 
%The Far Ultraviolet Spectroscopic Explorer has the wavelength coverage and sensitivity to detect \ovi{} lines in young sources with low visual extinctions like some of our targets.
An important parameter in the X--ray model is the input X--ray luminosity and 
spectral temperature which \citet{glass07} take from solar--like young stellar objects in Orion \citep{wolk05}. It will be important to explore the effect of different X--ray luminosities and spectra to better compare model predictions with observations. The anti--correlation between \neii{} luminosity and mass accretion rate (if confirmed by a larger sample of sources) could be used  to identify the dominant ionization mechanism for Ne atoms. 
Detections of \neii{} emission lines from classical TTSs, older TTSs, and young brown dwarfs will help sample the trend to mass accretion rates ranging from  $ \sim 10^{-8}$ down to $ \sim 10^{-10}$\,M$_\sun$\,yr$^{-1}$.

None of the optically thin systems we published in \citet{pascucci06} exhibits \neii{} emission at 12.81\,\micron . The majority of these sources have 3$\sigma$ line flux upper limits in the H$_2$\,S(2) line (that are representative for the \neii{} lines) about an order of magnitude lower than the \neii{} fluxes we detect in the four optically thick disks (see Table~4 in \citealt{pascucci06}). In addition, these optically thin systems span a wide range in X--ray luminosities that covers the X--ray luminosities of the sources with detected \neii{} emission. This indicates  that the optically thin sample in \citet{pascucci06} lack enough hot gas to emit detectable \neii{} emission. The EUV and X--ray models of optically thick dust disks as presented above 
can reproduce the observed \neii{} emission lines from only $\sim 6 \times 10^{-7}$\,$M_{\rm J}$ of gas within 10\,AU or $\sim 3 \times 10^{-5}$\,$M_{\rm J}$ of gas within 20\,AU, respectively. 
Thus, \neii{} non--detections in optically thin systems might be used to set even more stringent gas mass upper limits 
than those we reported in \citet{holl05} and \citet{pascucci06} using other mid--infrared gas transitions.
However, a correct determination of the gas mass upper limits requires detailed modeling of the star/disk properties. 
%an independent measurement of the stellar EUV flux to apply the EUV model. In the case of the X--ray model, it will be important to constrain the stellar X--ray flux as well as the disk properties. 
We plan to explore this issue in a separate contribution.

%In addition, the EUV and Xray models have different predictions on the strengths of
%other ionized lines, such as \siii{} 19\,\micron , which are weaker than
%     the \neii{} line.  More sensitive search for other infrared lines could
%     then help distinguish the origin of the \neii{} emission lines.

%In the case of EUV, the emission arises in 10$^4$ K fully ionized gas at the surface
%of the disk at $r < 10$ AU, with total ionized gas mass of $> 10^{-7}$\,$M_{\rm J}$.  In the case of X-rays, the emission
%arises in $\sim 10^3$\,K partially ionized gas (H$^+$/H and Ne$^+$/Ne $\sim 10^{-3} -- 10^{-2}$) at $r < 20$\,AU, with
%a total heated gas mass of $> 10^{-4}$\,$M_{\rm J}$.
%We can thus expect to find some correlations between the observed \neii{} luminosities and the disk/star properties, especially
%the X-ray and EUV luminosity of the stars.

\section{Summary}\label{S:summary}
To summarize, the main conclusions of this paper are as follows:
\begin{enumerate}
\item We detect \neii{} emission lines at 12.81\,\micron{} in four out of six optically thick dust disks observed as part of the FEPS {\it Spitzer} Legacy program.
The systems with \neii{} emission are characterized by weaker mid--infrared continuum (possibly the result of an inner 
hole and/or grain growth) compared to those systems where we do not detect \neii{} lines.
\item We also detect a \hi(7--6) emission line from RX~J1852.3-3700. Magnetospheric
accretion flows can account for at most $\sim$30\% of the observed flux. This line is not associated with the gas emitting the \neii{} lines. Accretion shocks and/or the stellar corona could contribute to most of the observed \hi(7--6) emission.
\item The \neii{} line luminosity correlates with the stellar X--ray luminosity. We find an anti--correlation between the \neii{} luminosity and the mass accretion rate. The slopes of these trends are not constrained by the current data.
\item Emission from Ne$^+$ is very likely arising from the hot surface of the disk. Both stellar X-rays and EUV photons can sufficiently ionize the disk surface to reproduce the observed line fluxes.
\end{enumerate}

\acknowledgments
It is a pleasure to thank all members of the FEPS team for their contributions to 
the project and to this study. IP wishes to thank D. Watson for suggestions in the data 
reduction of the IRS high--resolution spectra and A. E. Glassgold for helpful discussions on the 
X--ray model predictions. 
We thank the referee Dmitry Semenov
for a  very helpful review.
This work is based on observations made with 
the Spitzer Space Telescope, which is operated by the Jet Propulsion Laboratory, California 
Institute of Technology under NASA contract 1407. FEPS is pleased to acknowledge support through
NASA contracts 1224768, 1224634, and 1224566 administered through JPL. 

%% in parentheses, after the keyword, but they will not be verified.

{\it Facilities:} \facility{Spitzer Space Telescope}

\clearpage

\end{document}